\documentclass[prd,showpacs,preprintnumbers,amsmath,amssymb]{revtex4}
\usepackage{epsfig}
\usepackage[mathscr]{euscript}
\usepackage{mathtools}
\usepackage{graphicx}
\usepackage{float,epsfig}
\usepackage{dcolumn}% Align table columns on decimal point
\usepackage{bm}% bold math
\usepackage{graphicx}% Include figure files
\usepackage{amsmath,amssymb,amsthm}
\usepackage[colorlinks=true,linkcolor=blue]{hyperref}
\usepackage{booktabs}
\usepackage{dcolumn}% Align table columns on decimal point
\newcommand{\bea}{\begin{eqnarray}}
\newcommand{\eea}{\end{eqnarray}}
\newcommand{\beq}{\begin{equation}}
\newcommand{\eeq}{\end{equation}}

\def\/{\over}
\begin{document}
\title{Optimal estimation of parameters for scalar field in an expanding spacetime exhibiting Lorentz invariance violation}

\author{Xiaobao Liu}
\affiliation{Department of
Physics, Key Laboratory of Low Dimensional Quantum Structures and
Quantum Control of Ministry of Education, and Synergetic Innovation
Center for Quantum Effects and Applications, Hunan Normal
University, Changsha, Hunan 410081, P. R. China}

\author{Jiliang Jing\footnote{Corresponding author, Email: jljing@hunn.edu.cn}}
\affiliation{Department of
Physics, Key Laboratory of Low Dimensional Quantum Structures and
Quantum Control of Ministry of Education, and Synergetic Innovation
Center for Quantum Effects and Applications, Hunan Normal
University, Changsha, Hunan 410081, P. R. China}

\author{Jieci Wang}
\affiliation{Department of
Physics, Key Laboratory of Low Dimensional Quantum Structures and
Quantum Control of Ministry of Education, and Synergetic Innovation
Center for Quantum Effects and Applications, Hunan Normal
University, Changsha, Hunan 410081, P. R. China}

\author{Zehua Tian\footnote{Corresponding author, Email: tianzh@ustc.edu.cn}}
\affiliation{ CAS Key Laboratory of Microscale Magnetic Resonance and Department of Modern Physics, University of Science and Technology of China, Hefei 230026, China}

\affiliation{Hefei National Laboratory for Physical Sciences at the Microscale, University of Science and Technology of China, Hefei 230026, China}

\affiliation{Synergetic Innovation Center of Quantum Information and Quantum Physics, University of Science and Technology of China, Hefei 230026, China
}
\affiliation{Key Laboratory for Research in Galaxies and Cosmology,
Chinese Academy of Science, 96 JinZhai Road, Hefei 230026, Anhui, China}

\begin{abstract}
We address the optimal estimation of quantum parameters, in the framework of local quantum estimation theory, for a massive scalar quantum field in the expanding Robertson-Walker universe exhibiting Lorentz invariance violation (LIV). The information about the history of the expanding spacetime in the presence of LIV can be extracted by taking measurements on the entangled state of particle modes.
We find that, in the estimation of cosmological parameters, the ultimate bounds to the precision of the Lorentz-invariant massive scalar field can be improved due to the effects of LIV under some appropriate conditions.
We also show that, in the Lorentz-invariant massive scalar field and massless scalar field due to LIV backgrounds, the optimal precision can be achieved by choosing the particles with some suitable LIV, cosmological and field parameters. Moreover, in the estimation of LIV parameter during the spacetime expansion, we prove that the appropriate momentum mode of field particles and larger cosmological parameters can provide us a better precision. Particularly, the optimal precision of the parameters estimation can be obtained by performing projective measurements implemented by the projectors onto the eigenvectors of specific probe states.
\end{abstract}
\keywords{Lorentz invariance violation; expanding universe; local quantum estimation theory}
\pacs{04.60.-m, 04.62.+v, 06.20.-f}

\maketitle

\section{introduction}
The theory of general relativity plays a dominant role in the field with a very high density, where the classical physics breaks down and quantum gravity may be important to resolve the big-bang singularity
which is the prediction of classical general relativity.
Until now there have many different quantum gravity approaches, such as string theory~\cite{Green,Polchinski} and loop quantum gravity~\cite{Rovelli,Thiemann}.
However, several approaches to quantum gravity suggest that a microscopic structure of spacetime may lead to the Lorentz symmetry violation. In the last decades, this symmetry violation has been investigated in the context of noncommutative geometry~\cite{Connes0,Connes1,Majid},  the extra dimensions~\cite{Overduin,Rizzo}, and the discretization process. Besides, there has a lot of physical phenomena showing that the Lorenz symmetry may be broken~\cite{Greisen,Takeda,Krennrich,Antonov,Sato,Coleman,Zhang2018}.
It is well known that the standard dispersion relation
represents the relationship between the energy and momentum of a particle, which is invariant under continuous Lorentz transformations. However,
when the distances are as small as the Planck length, the spacetime might have a discrete structure and quantum gravity effects are dominant.
This is the smallest length scale at which the standard dispersion
relation should be modified by adding an extra term that violates the invariance of Lorentz symmetry~\cite{Preparata0,Cacciatori,Preparata1,Xue,Zarei,Motie,Jacobson,Liberati,Mattingly}. Recently,
the effects of LIV have not only been studied in inflationary cosmology~\cite{Brandenberger,Easther,Starobinsky,Martin,Khajeh,Mohammadzadeh}, but also shown that it is relevant to the dark energy problem and baryogenesis~\cite{Bertolami0,Bertolami1,Alberghi}.

On the other hand, many quantities of interest to us correspond to nonlinear functions of the density operator and thus cannot be associated with quantum observables. As a consequence, any procedure aims at evaluating the quantity of interest which is ultimately a parameter estimation problem, where the value of the quantity of interest is indirectly inferred from the measurement of one or more proper observables. Therefore, an optimization problem naturally arises, which may be properly addressed in the framework of local quantum estimation theory (QET)~\cite{Helstrom, Holevo, Braunstein, Pairs, Cramer}, providing a analytical tool to find the optimal measurement according to some given criterion. Relevant examples of this application of QET are given by discussions of gravitational wave detection~\cite{Vallisneri,Aasi}, measurements of entanglement~\cite{Genoni,Brida}, Unruh-Hawking temperature and phase parameters for accelerated detectors~\cite{Aspachs2010,Hosler2013,Yao2014,Wang2014,Tian2015,Yang2018}, deformed parameter for the  noncommutative spacetime~\cite{Liu2018}, and so on. Recently, with the help of local QET, Wang \emph{et al}. have investigated the ultimate precision limits for the estimation of expansion parameters of universe~\cite{Wang2015}, while such technology has not yet been applied to the quantum parameter estimation due to the effects of LIV on the expanding Robertson-Walker (RW) universe.

In this paper, we investigate the ultimate limits of precision in the estimation of parameters, for a massive scalar field on a two-dimensional conformally flat RW expanding spacetime in the presence of LIV, employing the local QET.
It is worth noting that particles can be created by propagation of a massive scalar field due to the expansion of a RW spacetime~\cite{Birrell,Ball,Zehua}.
Moreover, as considered in Refs.~\cite{Khajeh,Mohammadzadeh}, the effects of LIV on the massless scalar field in an expanding universe can also lead to the creation of particles.
Otherwise, the spacetime is conformally equivalent to Minkowski spacetime, and the conformally invariant massless scalar field does not produce particles in the absence of LIV.
Therefore, we will explicitly calculate the quantum Fisher information (QFI), i.e., derive the ultimate bounds to the precision in the estimation of LIV coefficient and cosmological parameters.
By performing projective measurements implemented by the projectors onto the eigenvectors of specific probe states, we find that the Fisher information (FI) with respect
to the population measurement on the final particles state is equal to that of QFI. Therefore, this measurement is the optimal measurement in the estimation of parameters.
Especially, due to the effects of LIV, the bounds on the measurement of cosmological parameters for the Lorentz-invariant massive scalar field case are possible to be enhanced.
In addition, for the estimation of LIV parameter, the precision can be improved by choosing the projective measurements corresponding to eigenvectors of the particle states with appropriate LIV, cosmological and field parameters.

The paper is organized as follows. In Sec. II we introduce the quantization of massive scalar field on a RW spacetime due to the effects of LIV.
In Sec. III, we briefly take a review of the basic notions of local quantum estimation theory.
In Sec. IV, we calculate the QFI in the estimation of cosmological parameters, and study how does the existence of LIV modify what the bounds on measurement for the Lorentz-invariant massive scalar field case. We also discuss
the QFI in two specific cases, i.e., the massive scalar field in the absence of LIV and the massless scalar field due to LIV during the evolution of expansion spacetime.
In Sec. V, we investigate the QFI in the estimation of the LIV parameter to find the highest precision for the parameters estimation.
Finally, a summary of the main results of our work is present in Sec. V.

%%%%%%%%%%%%%%%%%%%%%%%%%%%%%%%
\section{the cosmological model in the presence of Lorentz invariance violation}
Let us start with a two-dimensional Robertson-Walker expanding spacetime with the line element
\begin{eqnarray}\label{line element}
ds^2=dt^2-a^2(t)dx^2,
\end{eqnarray}
where $a(t)$ is the scale factor. By introducing the conformal time $\eta$ which relates to cosmological time $t$ by $d\eta=a^{-1}(t)dt$, the metric in Eq.~(\ref{line element}) takes the form
\begin{eqnarray}\label{metric form}
ds^2=c^2(\eta)(d\eta^2-dx^2),
\end{eqnarray}
where $c^2(\eta)=a^2(t(\eta))$ is the conformal scale factor. As considered in the Refs.~\cite{Khajeh,Mohammadzadeh}, the generalized Lagrangian for the cosmological spacetime in the presence of LIV can be written as
\begin{eqnarray}\label{Lagrangian}
\mathcal{L}=\frac{1}{2}\sqrt{-g}[g^{\mu\nu}\nabla_\mu \phi\nabla_\nu \phi-m^2\phi^2+\alpha^2(D^2\phi)^2
+\lambda(1-\mu^{\mu}\mu_\mu)],\nonumber\\
\end{eqnarray}
where $m$ is mass of the scalar field particles, $\alpha^2$ denotes the LIV coefficient. The covariant spatial Laplacian is given by
\begin{eqnarray}\label{spatial Lagrangian}
D^2\phi=-D^{\mu}D_{\mu}\phi=-q^{\mu\nu}\nabla_\nu(q^\tau_\mu\nabla_\tau\phi).
\end{eqnarray}
where $q_{\mu\nu}=-g_{\mu\nu}+\mu_{\mu}\mu_\nu$ with $g_{\mu\nu}$ corresponding to the metric in Eq.~(\ref{line element}). The vector field $\mu_{\mu}$ is as a non-dynamical vector field to be specified by the conditions of the theory.
The Lagrange multiplier $\lambda$ in Eq.~(\ref{Lagrangian}) constrains  $\mu^{\mu}$ as
\begin{eqnarray}\label{constrains}
g_{\mu\nu}\mu^{\mu}\mu^\nu=1.
\end{eqnarray}
Due to the Robertson-Walker metric describing a homogeneous and isotopic spacetime, the vector field $\mu_\mu$ is taken as to satisfy the isotropic property of cosmological metric and the Eq.~(\ref{constrains}).
Thus, we have $\mu^\mu=(c(\eta),0)$.
By equating the variation of the action $S=\int \mathcal{L}d^4x$ with respect to $\phi$ to zero,
the equation of motion for scalar field can be obtained as follow
\begin{eqnarray}\label{motion equation}
\square\phi-m^2c^2(\eta)-\frac{\alpha^2}{c^2(\eta)}\partial_x^4\phi=0,
\end{eqnarray}
where $\square:=\frac{1}{\sqrt{-g}}\partial_\mu(\sqrt{-g}g^{\mu\nu}\partial_\nu)$ is the d'Alembertian operator,
and the conformal factor is~\cite{Birrell,Ball,Bernard,Duncan,Fuentes,Martin0,Moradi}
\begin{eqnarray}\label{conformal scale factor}
c^2(\eta)=1+\epsilon\left[1+\tanh(\rho\eta)\right],
\end{eqnarray}
with $\epsilon$ and $\rho$ being positive real parameters corresponding to the total volume and rate of the expansion for the universe, respectively.
In the far past and future, the spacetime becomes Minkowskian spacetime, due to the fact that $c^2(\eta)\rightarrow 1$ in the asymptotic \emph{in}-region $(\eta\rightarrow -\infty)$ and $c^2(\eta)\rightarrow 1+2\epsilon$ in the asymptotic \emph{out}-region $(\eta\rightarrow +\infty)$.

To find the solutions to the scalar field equation of motion, we separate it into positive and negative frequency modes, i.e., $u_k(\eta,x)$ and $u_k^\ast(\eta,x)$.
By invoking the method of separation of variables, we have the mode solution  $u_k(\eta,x)=e^{ikx}\chi_k(\eta)$, such that $\chi_k(\eta)$ satisfies
\begin{eqnarray}\label{mode solution}
\frac{d^2}{d\eta^2}\chi_k(\eta)+\bigg[k^2+m^2c^2(\eta)-\frac{\alpha^2}{c^2(\eta)}k^4\bigg]\chi_k(\eta)=0.
\end{eqnarray}
Note that the dispersion relation is modified by adding a LIV term which violates the invariance of Lorentz symmetry due to the quantum gravity effects~\cite{Amelino}.
Thus, the modified dispersion relation in the expanding spacetime is
\begin{eqnarray}\label{modified dispersion relation}
\omega^2 =k^2+m^2c^2(\eta)-\frac{\alpha^2}{c^2(\eta)} k^4.
\end{eqnarray}
It is worth mentioning that the LIV coefficient $\alpha^2$ is proportional to the Planck length
$l_p$ which gives the semiclassical limit beyond the standard model, such that the result recover to the
dispersion relation in an expanding universe if we take $l_p\rightarrow 0$.
%in this paper, we only consider the massless scalar field case, i.e., $m=0$,
On the other hand,  the equation (\ref{mode solution}) can be solved by the hypergeometric functions.
In the asymptotic \emph{in}-region, the normalized modes are found to be
\begin{eqnarray}\label{mode1}
u_{k}^{\rm in}(\eta,x)&=&\frac{1}{\sqrt{4\pi\omega_{\rm in}}}\nonumber\\
&\times&
\exp\bigg(i k x-i\omega^+_k\eta-i\frac{\omega^-_k}{\rho}\ln[(1+2\epsilon)e^{\rho\eta}+e^{-\rho\eta}]\bigg)\nonumber\\
&\times&F(1+i\omega^-_k/\rho,i\omega^-_k/\rho;1-i\omega_{\rm in}/\rho;z),
\end{eqnarray}
and in the asymptotic \emph{out}-regions, the normalized modes are
\begin{eqnarray}\label{mode2}
u_{k}^{\rm out}(\eta,x)&=&\frac{1}{\sqrt{4\pi\omega_{\rm in}}}\nonumber\\
&\times&
\exp\bigg(i k x-i\omega^+_k\eta-i\frac{\omega^-_k}{\rho}\ln[(1+2\epsilon)e^{\rho\eta}+e^{-\rho\eta}]\bigg)\nonumber\\
&\times& F(1+i\omega^-_k/\rho,i\omega^-_k/\rho;1-i\omega_{\rm out}/\rho;z)\;,
\end{eqnarray}
with $F$ being the ordinary hypergeometric function and
\begin{eqnarray}
&&\omega_{\rm in}=\sqrt{k^2+m^2-\alpha^2k^4}, \nonumber\\
&&\omega_{\rm out}=\sqrt{k^2+m^2(1+2\epsilon)-\frac{\alpha^2}{1+2\epsilon}k^4}, \nonumber\\
&&\omega_k^\pm=\frac{1}{2}(\omega_{\rm in}\pm\omega_{\rm out}), \nonumber\\
&&z=\frac{1+2\epsilon}{2}\frac{1+\tanh(\rho\eta)}{1+\epsilon\tanh(\rho\eta)}.
\end{eqnarray}

Now it is possible to quantize the field and obtain
the relation between the vacuum state in the far past region and that in the far future region.
Using the inner product and the linear transformation properties of hypergeometric function, the Bogoliubov transformations associated with the \emph{in} and \emph{out} solutions take the form
\begin{eqnarray}\label{transformations}
u_k^{\rm in}(\eta,x)=\alpha_k u_k^{\rm out}(\eta,x)+\beta_k u_{-k}^{\rm out}(\eta,x),
\end{eqnarray}
where the Bogoliubov coefficients $\alpha_k$ and $\beta_k$ are given by
\begin{eqnarray}\label{Bogoliubov coefficients}
\alpha_k &=&\sqrt{\frac{\omega_{\rm out}}{\omega_{\rm in}}}
\frac{\Gamma(1-i\omega_{\rm in}/\rho)\Gamma(-i\omega_{\rm out}/\rho)}
{\Gamma(-i\omega_{k}^+/\rho)\Gamma(1-i\omega_{k}^+/\rho)},\nonumber\\
\beta_k &=&\sqrt{\frac{\omega_{\rm out}}{\omega_{\rm in}}}
\frac{\Gamma(1-i\omega_{\rm in}/\rho)\Gamma(i\omega_{\rm out}/\rho)}
{\Gamma(-i\omega_{k}^-/\rho)\Gamma(1-i\omega_{k}^-/\rho)}.
\end{eqnarray}
To find the relation between the far past and future vacuum states, we use the relationship between the operators
$b_{\rm in}(k)=\alpha^*b_{\rm out}(k)-\beta^*b_{\rm out}^\dagger(k)$.
Here, the creation and annihilation operators, $b_{\rm in/out}(k)$ and $b_{\rm in/out}^\dagger(k)$, satisfy the commutation
relations $[b_{\rm in}(k),b_{\rm in}^\dagger(k')]=\delta_{kk'}$ and  $[b_{\rm out}(k),b_{\rm out}^\dagger(k')]=\delta_{kk'}$.
According to the definition of vacuum state $b_{\rm in}(k)|0\rangle_{\rm in}=0$ and the normalization condition $_{\rm in}\langle0|0\rangle_{\rm in}=1$, we can express the input vacuum state in terms of
the bosonic Fock basis in the \emph{out}-regions, which is
\begin{eqnarray}\label{vacuum state}
|0\rangle_{\rm in}=\frac{1}{|\alpha_k|}\sum_{n=0}^{\infty}\bigg(\frac{\beta^*_k}{\alpha^*_k}\bigg)^n
|n_k n_{-k}\rangle_{\rm out}.
\end{eqnarray}
When $\alpha^2=0$, i.e., an expanding universe is in the absence of LIV, the above
results recover to that obtained in Ref.~\cite{Birrell,Ball,Zehua} for a massive scalar field on a two-dimensional RW expanding spacetime as expected.
The authors demonstrated that particles are created by propagation of a massive quantum scalar field through an expanding universe,
while for the case of a massless quantum field, it does not produce particles due to the fact that the spacetime is conformally equivalent to Minkowski spacetime.
However, if the LIV occurs for massless scalar field in an expanding universe, i.e., $m=0$, the particles still can be generated due to the effects of LIV.  This is just
what has been shown in Ref.~\cite{Khajeh,Mohammadzadeh}.
Therefore, a single qubit prepared in a initial vacuum state $|0\rangle_{in}$ is converted to an entangled state in the particle number degree of freedom in the expanding spacetime.
We can extract the information about the LIV, cosmological, and field parameters which are codified in the final state.

%%%%%%%%%%%%%%%%%%%%%%%%%%%%%%%
\section{local quantum estimation theory}
Supposing that we have a given quantum state $\varrho(\lambda)$ parametrized by an unknown parameter $\lambda$, the unknown parameter may does not correspond to a proper quantum observable and it cannot be measured directly. To get information about $\lambda$, we have to resort to indirect measurements, inferring its value by the measurements of a set of observables. That is to say, we have a parameter estimation problem.
In the estimation problem we try to infer the value of a parameter $\lambda$  by measuring a different quantity $X$. Therefore,
any inference strategy amounts to find an estimator, i.e., a mapping $\hat{\lambda}=\hat{\lambda}(x_1,x_2,...,x_n)$ from the set of measurement outcomes into the space of parameters.
Optimal estimators are those saturating the Cram\'{e}r-Rao theorem~\cite{Cramer}
\begin{equation}\label{inequality}
\hbox{Var}(\lambda) \geq \frac{1}{ M F(\lambda)}\;,
\end{equation}
which establishes a lower bound on the variance $\hbox{Var}(\lambda)$ of any unbiased estimator of the parameter $\lambda$. Here, $M$ is
the number of measurements  and $F(\lambda)$ denotes the Fisher information (FI) which is
\begin{equation}\label{FI}
F(\lambda)=\sum_{x}p(x\mid \lambda)[\partial_{\lambda}\ln p(x\mid \lambda)]^2,
\end{equation}
where $\partial_{\lambda}=\frac{\partial}{\partial\lambda}$ and $p(x\mid\lambda)$ denotes the conditional probability corresponding measurement outcome $x$
with respect to a chosen positive operator valued measurement (POVM). According to the Born rule, we find $p(x\mid\lambda)=\hbox{Tr}[\rho(\lambda)\:\Pi_x]$, where $\{\Pi_x\}$ is the elements of POVM and $\rho(\lambda)$ denotes the density operator. Moreover,
we can maximize the FI over all the possible
quantum measurements on the quantum system. By introducing the symmetric Logarithmic derivative $L(\lambda)$ satisfying the partial differential equation
$\partial_{\lambda} \rho(\lambda)=  (\rho(\lambda)L(\lambda)+L(\lambda)\rho(\lambda))/2$, we can get that
the FI $F(\lambda)$ is upper bounded by the QFI $H(\lambda)$, i.e., $F(\lambda) \leq H(\lambda) \equiv \hbox{Tr}[\rho(\lambda)\:L(\lambda)^2]$. Therefore, we have the quantum Cram\'{e}r-Rao bound
\begin{equation}\label{inequality2}
\hbox{Var}(\lambda) \geq \frac{1}{M F(\lambda)} \geq \frac{1}{M H(\lambda)}\;,
\end{equation}
for the variance of any estimator, which represents that we have the ultimate bound to precision for any quantum measurement aimed at estimating the parameter $\lambda$ for a state of the family $\rho(\lambda)$.
Upon diagonalizing the density matrix as $\rho_{\lambda}=\sum_{i=1}^N p_i|\psi_i\rangle\langle\psi_i|$ with $p_i\geq 0$ and $\sum_{i=1} p_i=1$, the QFI can be rephrased as
\begin{equation}\label{SLD}
H(\lambda)=2\sum_{m,n}^N\frac{|\langle\psi_m|
\partial_{\lambda}\rho_{\lambda}|\psi_n\rangle|^2}{p_m + p_n}.
\end{equation}
For a quantum state with non-full-rank density matrix, the detailed formula for the QFI is~\cite{Pairs}
\begin{eqnarray}\label{QFI}
H(\lambda)=\sum_n \frac{(\partial_\lambda p_n)^2}{p_n}+
2\sum_{n,m}\frac{(p_n-p_m)^2}{p_n+p_m}|\langle\psi_m|
\partial_{\lambda}\psi_n\rangle|^2,
\end{eqnarray}
where the sums include only the terms with $p_n\neq 0$ and $p_n+p_m\neq0$.

%%%%%%%%%%%%%%%%%%%%%%%%%%%%%%%
\section{QFI for cosmological parameters estimation}
In this section, our aim is to discuss the effects of background structure on the precision of cosmological parameters estimation that appear in an expanding universe. Firstly, the cosmological parameter that we wish to estimate is the expansion volume $\epsilon$, based on the quantum Cram$\mathrm{\acute{e}}$r-Rao inequality~\cite{Cramer}.
It is worth mentioning that
the information about the total volume $\epsilon$ is encoded in the Bogoliubov coefficients. Hence we can  calculate the operational QFI in the estimation of expansion volume $\epsilon$ in terms of the Bogoliubov coefficients shown in Eq.~(\ref{Bogoliubov coefficients}).
Note that the reduced density matrix corresponding to particle states of the modes $k$ can be obtained by tracing out the modes $-k$ in the state (\ref{vacuum state}), which is
\begin{eqnarray}\label{particle state}
\varrho_k=\frac{1}{|\alpha_k|^2}\sum_{n=0}^{\infty}\bigg(\frac{|\beta_k|}{|\alpha_k|}\bigg)^{2n}
|n_k\rangle\langle n_k|.
\end{eqnarray}
Assuming that we are living in the spacetime which corresponds to the asymptotically flat regions of the future times, the volume of the expanding universe can be estimated by taking measurements on the particle state in Eq.~(\ref{particle state}).

Any estimation scheme can be performed in the following steps: (i) probe state prepared, (ii) interacting with a dynamic system in which the estimated parameters are involved,
(iii) measure the probe. In our paper, we consider the state of the particles as the probe.
Our task is to achieve the ultimate limit of precision for the expansion volume parameter $\epsilon$ estimation in the curved spacetime which is determined by the QFI.
However, in the quantum estimation process, we should carry out the optimization over measurement processes, i.e.,  maximizing the FI over all possible quantum measurements on the quantum system to obtain the QFI.
According to the expression of the QFI in Eq.~(\ref{QFI}), we have
\begin{eqnarray}\label{volume1}
H(\epsilon)=\sum_{n=0}^\infty\frac{(\partial_\epsilon\lambda_n)^2}{\lambda_n},
\end{eqnarray}
where $\lambda_n=\gamma^n-\gamma^{n+1}$ is the eigenvalues of the particle states of the probe with
\begin{eqnarray}
\gamma=\frac{|\beta_k|^2}{|\alpha_k|^2}=\frac{\sinh[\pi\omega_k^{-}/\rho]}
{\sinh[\pi\omega_k^{+}/\rho]}.
\end{eqnarray}

Moreover, we can obtain different FI in terms of the classical probabilities $p(x\mid\lambda)$ with respect to different POVM $\{\Pi_x\}$.
Since the QFI does not depend on any measurements, so it is hard to look for which measurement is optimal to obtain the ultimate bound.
However, we can maximize the FI over all possible quantum measurements on the quantum system to obtain the QFI.
Thus, to obtain the optimal measurement to estimate the expansion volume parameter, we can calculate the FI for the population measurement and then compare it with the QFI,
due to the fact that the central task is to determine whether the population measurement is optimal according to the condition of optimal quantum measurement, i.e., POVM with a FI equal to the QFI.
According to Eqs.~\eqref{FI} and ~\eqref{particle state}, we can obtain the FI for the population measurement which is
\begin{eqnarray}\label{volume2}
F(\epsilon)&=&\sum_{n=0}^\infty\frac{(\partial_{\epsilon}\lambda_n)^2}{\lambda_n}.
\end{eqnarray}
We are interested in finding that the FI $F(\epsilon)$ in Eq.~\eqref{volume2} equals to the QFI $H(\epsilon)$ in Eq.~\eqref{volume1}.
We know that the FI of any measurement process is upper bounded by the QFI, which means that the estimation of expansion volume parameter via the population measurement is the optimization over measurement processes now,
and then it is easy to find out the ultimate bound to precision of estimation of the expansion volume.
Therefore, we choose the projective measurement which is constituted by the eigenvector $|\psi_m\rangle$ of the final state, and then the eigenvalues of the final state are the measured probabilities.

\begin{widetext}

\begin{figure}[htbp]
\centering
\includegraphics[height=1.8in,width=2.5in]{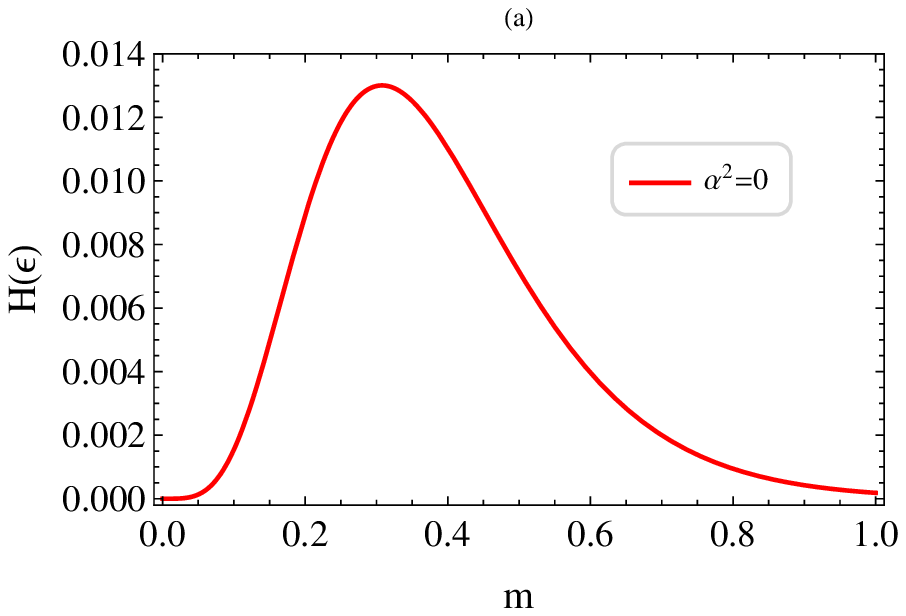}
\includegraphics[height=1.8in,width=2.5in]{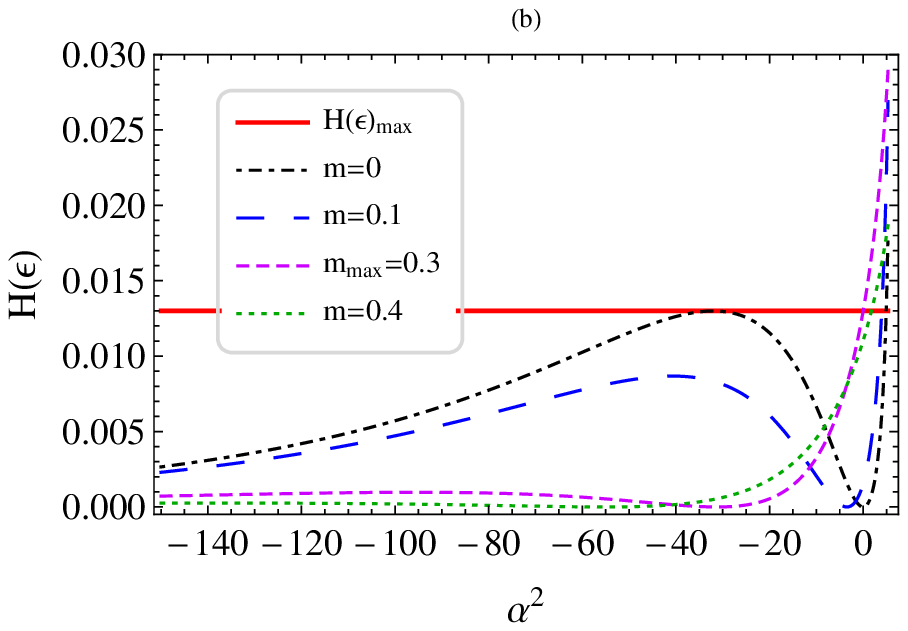}
\caption{ The parameters are fixed with $\epsilon=0.99$, $\rho=1$ and $k=0.4$. (a) For the Lorentz-invariant massive scalar case, the QFI $H(\epsilon)$ in terms of mass, $m$, of the scalar particles;
(b) For the case of massive scalar field in an expanding spacetime exhibiting LIV, the QFI $H(\epsilon)$ as a function of the LIV parameter, $\alpha^2$, with different values of mass, i.e., $m=0$ (dotted-dashed line), $m=0.1$ (large dashed line), $m_{\rm max}=0.3$ (dashed line) and $m=0.4$ (dotted line). Note that the maximum QFI $H(\epsilon)_{\rm max}$ (solid line) at the optimal mass corresponds to the Lorentz-invariant massive scalar case, for fixed $\alpha^2=0$ and $m_{\rm max}=0.3$.  This plot is not shown for some regions where $\omega_{\rm in}$ becomes imaginary. }\label{Figm1}
\end{figure}

\begin{figure}[htbp]
\centering
\includegraphics[height=1.8in,width=2.5in]{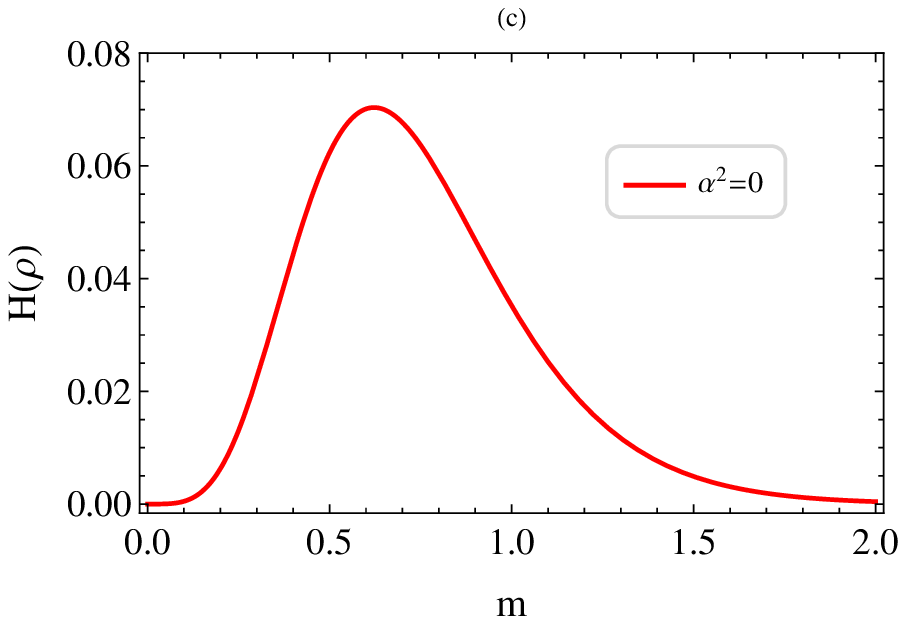}
\includegraphics[height=1.8in,width=2.5in]{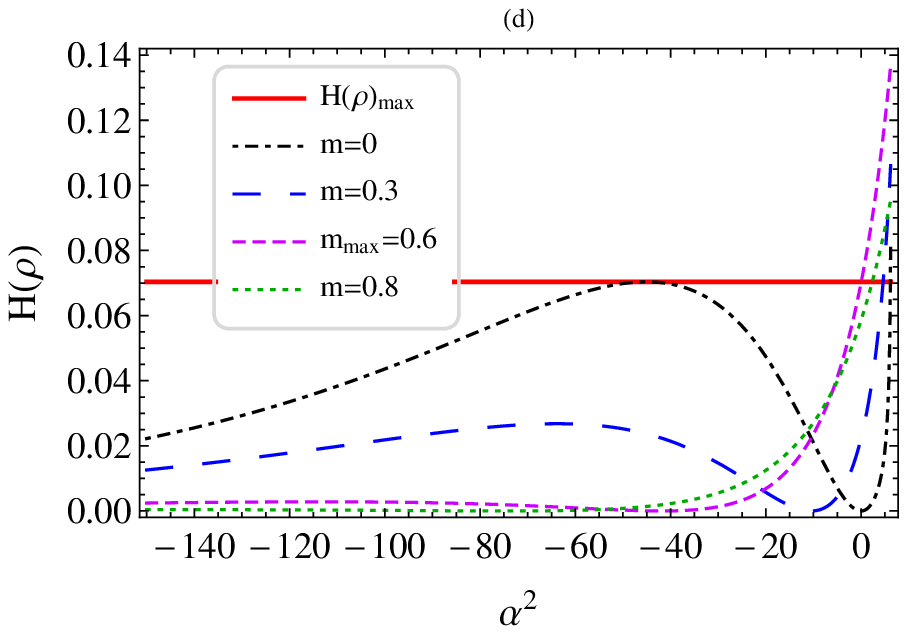}
\caption{The parameters are fixed with $\rho=1$, $\epsilon=0.99$ and $k=0.4$. (c) For the Lorentz-invariant massive scalar case, the QFI $H(\rho)$ over the mass, $m$, of the scalar particles;
(d) For the case of massive scalar field in an expanding spacetime exhibiting LIV, the QFI $H(\rho)$ with respects to the LIV parameter, $\alpha^2$, with different values of mass, i.e., $m=0$ (dotted-dashed line), $m=0.3$ (large dashed line), $m_{\rm max}=0.6$ (dashed line) and $m=0.8$ (dotted line). Note that the maximum QFI $H(\rho)_{\rm max}$ (solid line) at the optimal mass corresponds to the Lorentz-invariant massive scalar case, for fixed $\alpha^2=0$ and $m_{\rm max}=0.6$.  This plot is not shown for some regions where $\omega_{\rm in}$ becomes imaginary. }\label{Figm2}
\end{figure}

\end{widetext}

As a starting point, when modes of a massive scalar quantum field in an expanding spacetime can create particles,
we plot the QFI in the estimation of expansion volume, $H(\epsilon)$, with respect to mass, $m$, in Fig.~\ref{Figm1}(a) without the effects of LIV, for fixed values $\epsilon=0.99$, $\rho=1$ and $k=0.4$.
It shows that the bound on the parameter measurement of expansion volume, $\epsilon$, in terms of mass has maxima, which implies the precision peaks at the optimal value of $m$ ($m_{\rm max}=0.3$).
In turn, to investigate how does the existence of LIV modify what the bounds on measurement would be from the case of massive scalar field in the absence of LIV,
Fig.~\ref{Figm1}(b) shows the QFI $H(\epsilon)$ as a function of the LIV parameter $\alpha^2$.
We note that there exists two different choices for $\alpha^2$ in this figure, a positive one and a negative one. For the positive values of $\alpha^2$, there is an upper bound for $k$, due to the fact that $\omega_{\rm in}$ is an imaginary number for the value of $k$ greater than the upper bound. Thus, there is a forbidden region that we have not shown in the plot.
It is obvious from the Fig.~\ref{Figm1}(b) that for the negative values of $\alpha^2$, the $\omega_{\rm in/out}$ is always valid. In this case, we find that the QFI under the effects of LIV is always lower than the maximum value of the QFI, $H(\epsilon)_{\rm max}$, corresponding to the case of massive scalar field on an expanding spacetime in the absence of LIV shown in Fig.~\ref{Figm1}(a).
However, for the valid positive values, there exists some regions where the QFI due to the effects of LIV is larger than $H(\epsilon)_{\rm max}$,
which means that, the bound of precision in estimating the expansion volume for the Lorentz-invariant ($\alpha^2=0$) massive scalar quantum field case can be improved by the effects of LIV during the evolution of curved spacetime.
Moreover, with increasing the value of mass from $0$ to $m_{\rm max}$, the critical point of $\alpha^2$, corresponding to the case where the QFI due to LIV is bigger than $H(\epsilon)_{\rm max}$, is shifted toward to the point of origin. While the mass of scalar particles increases from $m_{\rm max}$ to infinity, the critical point of $\alpha^2$ is shifted toward positive values.

Similar to the expansion volume $\epsilon$ that we have estimated, we will briefly discuss the bounds on measurement of expansion rate for two situations, i.e., the massive scalar field without the LIV and in the presence of LIV during the evolution of expanding spacetime, respectively.
With the help of Eqs.~\eqref{QFI} and ~\eqref{particle state}, it is easy to obtain the formula of the QFI in the estimation of expansion rate, and to find that the QFI depends on the parameters $k$, $m$, $\alpha^2$, $\epsilon$ and $\rho$. In Fig.~\ref{Figm2}(c), we plot the behavior of the QFI $H(\rho)$ as a function of mass, $m$, for the Lorentz-invariant massive scalar case, by taking $\rho=1$, $\epsilon=0.99$ and $k=0.4$. This plot shows that the QFI reaches to the upper bound at the optimal value of mass, i.e., $m_{\rm max}=0.6$, which means that the optimal precision for the expansion rate peaks at certain mass of particles.
However, since a dynamic LIV model can modify the bounds on the measurement of expansion rate for the Lorentz-invariant massive scalar case, we illustrate the QFI $H(\rho)$ in terms of the LIV parameter, $\alpha^2$, in Fig.~\ref{Figm2}(d). Similarly, for the negative values of $\alpha^2$, the QFI is always smaller than $H(\rho)_{\rm max}$ which is the maximum QFI of the Lorentz-invariant massive scalar case shown in Fig.~\ref{Figm2}(c).
For the valid positive values of $\alpha^2$, the QFI due to the effects of LIV is possible larger than $H(\rho)_{\rm max}$, which implies that the bound for precision in estimating the expansion rate for the Lorentz-invariant massive scalar case can be enhanced by the effects of LIV. In addition, when the value of mass is $m_{\rm max}=0.6$, the QFI under the effects of LIV is the first to larger than $H(\rho)_{\rm max}$.

Next, to investigate how precisely one can estimate the cosmological parameters, we want to discuss two specific cases: a massive scalar field in the absence of LIV and a massless scalar field exhibiting LIV during the evolution of expanding universe.  It is worth mentioning that the behaviors of the QFI in the estimation of expansion rate, $\rho$, are similar to that the QFI in the estimation of expansion volume, $\epsilon$, so we would focus on discussing the behavior of QFI in the estimation of expansion volume, $H(\epsilon)$, in detail.

\subsection{The case for $\alpha^2=0$}
We first consider the case for a massive scalar quantum field on a two-dimensional asymptotically flat RW expanding spacetime in the absence of LIV, i.e., $\alpha^2=0$.
Note that the QFI, $H(\epsilon)$, in Eq.~\eqref{volume1} depends on the parameters $m$, $k$, $\epsilon$ and $\rho$.
Now let us examine the behavior of the precision of the expansion volume estimation for the massive scalar field on the RW expanding spacetime without the effects of LIV.

\begin{figure}[htbp]
\centering
\includegraphics[height=1.8in,width=2.8in]{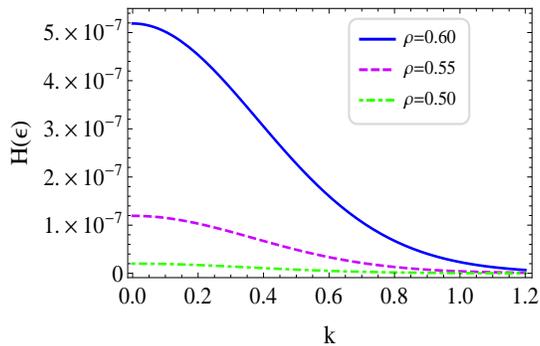}
\caption{ For the case of Lorentz-invariant massive scalar field in an expanding spacetime, the QFI $H(\epsilon)$ as a function of the momentum, $k$, with three values of expansion rate $\rho$, for fixed $\epsilon=0.99$ and $m=1$.}\label{Figurem1}
\end{figure}

Figure \ref{Figurem1} shows the behavior of the QFI, $H(\epsilon)$, with respect to momentum of the field mode $k$ with three different expansion rate, for fixed values $\epsilon=0.99$ and $m=1$.
It is obvious that the precision of the estimation of expansion volume is a monotonic decreasing function of the momentum of the mode. We are interested in finding that
the behavior of QFI in the estimation of expansion volume in this paper is very similar with the entanglement behavior for a massive scalar field in the expanding RW universe shown in Fig.~1 of Ref.~\cite{Fuentes}.
This implies that the precision of expansion volume estimation can be improved if entangled states are used as probe systems. We also see that the QFI is sensitive to the rate of expanding universe. As the expansion rate grows,
the precision of quantum parameter estimation increases, due to the increment of quantum entanglement.

\begin{figure}[htbp]
\centering
\includegraphics[height=1.8in,width=2.8in]{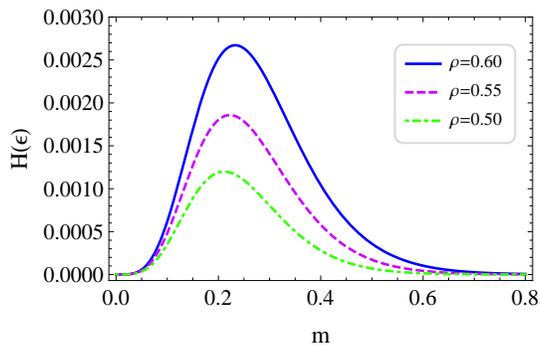}
\caption{(color online). For the case of a massive scalar quantum field in the expanding RW universe, QFI in the estimation of the expansion volume with respect to the mass $m$ of the scalar particles with three values of $\rho$. The parameters  are fixed $\epsilon=0.99$ and $k=0.4$.}\label{Figurem2}
\end{figure}

In Fig.~\ref{Figurem2}, we plot the behavior of QFI, $H(\epsilon)$, in the estimation of expansion volume as a function of mass $m$ of the scalar particles for different values of $\rho$.
It is shown that the QFI increases for a while and then starts to decrease as the increasing mass $m$. This is reminiscent
from the fact that for some very larger masses, the quantum entanglement diminishes \cite{Fuentes}, as a consequence of that the QFI disappears accordingly.
Therefore, we can arrive to a conclusion that the highest precision in the estimation of expansion volume can be arrived at a certain mass in the quantum measuring process.

\begin{figure}[htbp]
\centering
\includegraphics[height=1.8in,width=2.8in]{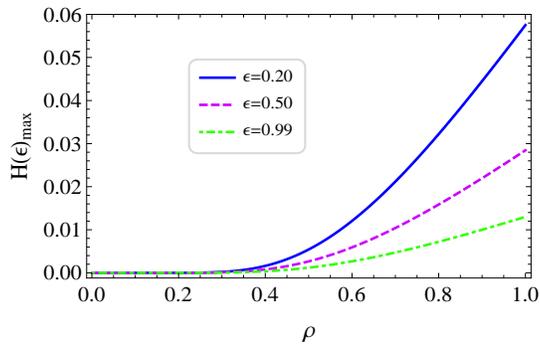}
\caption{For the case of a massive scalar quantum field in the expanding RW universe, QFI $H(\epsilon)$ at the optimal mass $m$ as a function of $\rho$ for different values of $\epsilon=0.20,0.50,0.99$. The parameter is fixed $k=0.4$.}\label{Figurem3}
\end{figure}

For the massive scalar field on a two-dimensional conformally flat RW universe in the absence of LIV, there is a privileged value of mass $m$ for which the QFI in the estimation of expansion volume has a optimal precision.
Therefore, in Fig.~\ref{Figurem3}, we present that, for different values of $\epsilon$, the maximum value of the QFI $H(\epsilon)$ at the optimal point of mass $m$ is a function of rate $\rho$ of the expansion.
Indeed, from this figure we can deduce that the precision of expansion volume estimation enhances as the expansion rate gets larger values. Also,
the smaller the expansion volume, the bigger the QFI is, which means that it is easier to achieve a given estimation precision with smaller expansion volume.

\subsection{The case for $m=0$}
When the massless scalar field equation in an expanding universe is modified by adding an extra term, it breaks the conformal invariance leading to the existence of LIV. Here we consider the case $m=0$.
The reason that we are interested is because LIV in an expanding universe can also lead to the creation of cosmological particle, even for a massless scalar fields in curved spacetime.
In this case, we can see that the QFI of the expansion volume parameter depends on $\alpha^2$, $k$, $\epsilon$ and $\rho$.

\begin{figure}[htbp]
\centering
\includegraphics[height=1.8in,width=2.8in]{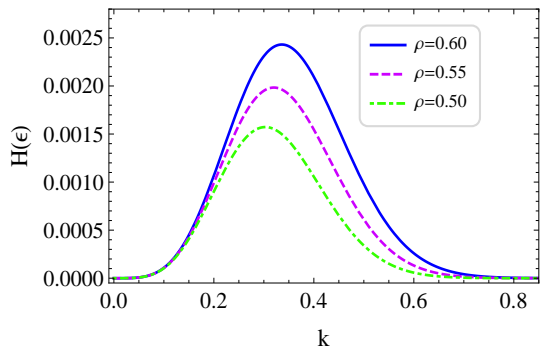}
\caption{For the case of massless scalar field in an expanding spacetime exhibiting LIV, the QFI $H(\epsilon)$ as a function of the momentum modes $k$, for fixed values $\epsilon=0.99$ and $\alpha^2=-10$,
and three values of parameter $\rho$, i.e., $\rho=0.60$ (solid line), $\rho=0.55$ (dashed line), $\rho=0.50$ (dotted-dashed line), respectively.}\label{Figure1}
\end{figure}
We plot the QFI in the estimation of expansion volume $\epsilon$, in Figure \ref{Figure1}, as a function of the momentum modes $k$ with three different values of expansion rate $\rho$, for fixed parameters $\epsilon=0.99$ and $\alpha^2=-10$. It is shown that the QFI firstly increases and then starts to decrease by increasing the value of $k$, which means that the highest precision in the expansion volume estimating can be obtained at a specific momentum mode.
This behavior of QFI is in analogy with the behavior of quantum entanglement for a massless scalar field in the presence of LIV on a two-dimensional asymptotically flat RW expanding spacetime~\cite{Khajeh}, which indicates that the precision of expansion volume estimation will be improved by increasing of quantum entanglement.
Moreover, as the expansion rate, $\rho$, grows, the QFI increases accordingly. Therefore, the higher precision in the estimation of expansion volume can be achieved by increasing the expansion rate.

%the QFI of the probe state sensitively depends on momentum $k$ of the scalar field particles.

\begin{figure}[htbp]
\centering
\includegraphics[height=1.8in,width=2.8in]{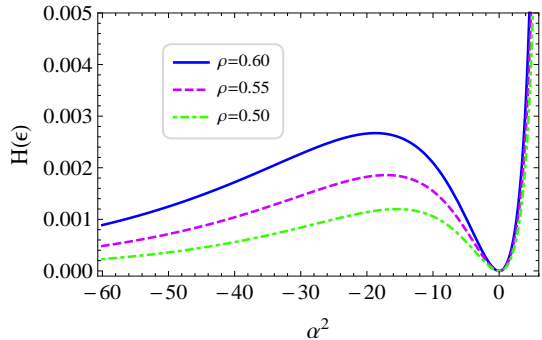}
\caption{For the case of massless scalar field in an expanding spacetime exhibiting LIV, the QFI $H(\epsilon)$ in terms of the LIV parameter, $\alpha^2$, for fixed values $\epsilon=0.99$ and $k=0.4$, with three different values of $\rho$, i.e., $\rho=0.60$ (solid line), $\rho=0.55$ (dashed line),and $\rho=0.50$ (dotted-dashed line). Note that the plot is not shown for some regions where $\omega_{\rm in}$ becomes imaginary.}\label{Figure2}
\end{figure}
Figure \ref{Figure2} shows the QFI in the estimation of expansion volume with respect to the LIV parameter, $\alpha^2$, with different values of expansion rate $\rho$.
We note that there is a forbidden region that we have not shown in the plot. For the valid positive values of $\alpha^2$, it is obvious from the plot that the precision of the estimation of expansion volume $\epsilon$ will increase monotonically by increment in the positive value of $\alpha^2$. However, for the negative one, we find that the QFI increases for a while for a specific range of absolute values of $\alpha^2$ and then decreases, which means that the precision of estimation has the maximum value at a certain value of LIV parameter $\alpha^2_{\rm max}$. In addition, for the case $\alpha^2=0$, the cosmological particles cannot be created due to no difference between the asymptotically flat regions in the past and future times, i.e., $\omega_{\rm in}=\omega_{\rm out}$. Therefore, no one can extract any information from the expansion spacetime, which means that the parameter estimation has the lowest precision when $\alpha^2=0$ as expected. It is worth noting that the behavior of quantum entanglement for a massless scalar field due to LIV during the evolution of expanding spacetime shown in Ref. \cite{Khajeh}, is similar with the behavior of QFI with respect to $\alpha^2$ in our paper, which represents that the precision of expansion volume estimation will be enhanced by the increment of quantum entanglement.

\begin{figure}[htbp]
\centering
\includegraphics[height=1.8in,width=2.8in]{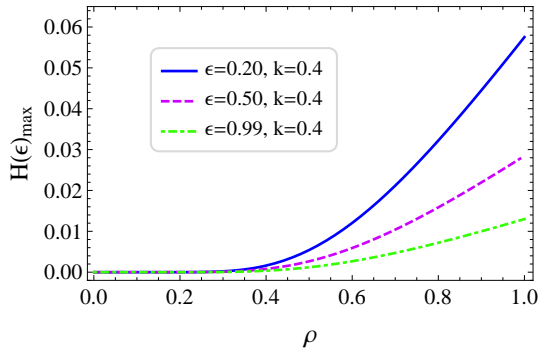}
\caption{ The maximum QFI in the optimal value of momentum mode $k$ as a function of expansion rate $\rho$ for fixed value of momentum mode $k=0.4$, and three different values of expansion volume $\epsilon=0.20, 0.50, 0.99$ (solid, dashed, dotted-dashed).}\label{Figure4}
\end{figure}

We benefit from figure \ref{Figure1} that with a any set of given values of $\alpha^2$, $\epsilon$ and $\rho$, QFI reaches a maximum value at a certain momentum $k$.
In turn, to investigate how does the choice of cosmological parameters which are allowed for the expanding universe affect this maximum value of QFI, $H(\epsilon)_{\rm max}$, at the optimal point of $k$, i.e., for fixed optimal value $k=0.4$, we plot the maximum value of QFI at the optimal point of momentum mode as a function of expansion rate in Fig.~\ref{Figure4}. Three different expansion volume $\epsilon$ have shown in this figure. For each given expansion volume $\epsilon$, the maximum QFI increases by the increment in the rapidity of the expanding universe, which means that the precision for the volume estimation can be enhanced by a larger expansion rate. Furthermore, the maximum QFI increases as the expansion volume gets smaller values, which indicates that it is easier to achieve a given precision for the expansion volume estimation.

\begin{figure}[htbp]
\centering
\includegraphics[height=1.8in,width=2.8in]{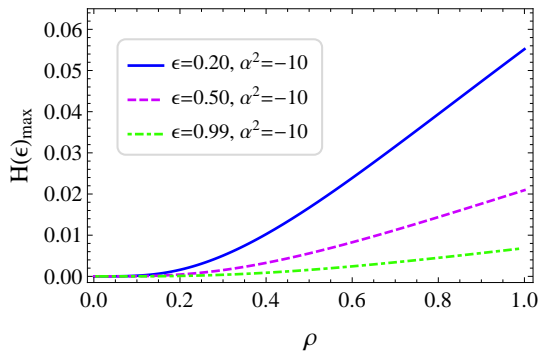}
\caption{ QFI in the optimal LIV parameter $\alpha^2$ as a function of expansion rate $\rho$ for fixed value of LIV parameter $\alpha^2=-10$ and three different values of expansion volume $\epsilon=0.20, 0.50, 0.99$  (solid, dashed, dotted-dashed).}\label{Figure5}
\end{figure}
Similarly, to study how does the choice of cosmological parameters which are allowed for the expanding universe affect this maximum value of QFI for the fixed optimal value $\alpha^2=-10$,
in Fig.~\ref{Figure5}, we plot the maximum value of the QFI at the optimal point of LIV parameter, $\alpha^2$, with respect to the expansion rate and three different expansion volume. We find that as the expansion rate grows, the maximum QFI for the optimal value of $\alpha^2$ increases accordingly. Therefore, the precision of estimation of expansion volume is an increasing function of the expansion rate. In addition, as can be seen, the smaller value of $\epsilon$, the bigger QFI, i.e., the easier to obtain the highest precision in the estimation process.

%%%%%%%%%%%%%%%%%%%%%%%%%%%%%%%
\section{QFI for LIV parameter estimation during the spacetime expansion}
To investigate how precisely we can estimate the LIV parameter $\alpha^2$ which modifies the dispersion relation in an expanding spacetime, we suppose the massless case in this section, i.e., $m=0$, without loss of generality.
This is reduced to a parameter estimation problem based on the Cram\'{e}r-Rao inequality~\cite{Cramer}.
Note that pairs of entangled particles are generated in the final states of field, then the information of LIV parameter is encoded in the Bogoliubov coefficients.
Assuming that we are living in the spacetime which corresponds to the asymptotically flat regions of the future times, we can now estimate the LIV parameter by taking measurements on the particle state in Eq.~(\ref{particle state}).

The key issue in estimation of LIV parameter is to find the optimal probe state preparation and related parameters that allow us to obtain the largest QFI.
In this system, the state of the particles act as the probe. Physically, a fixed value of FI can be obtained by any set of measurement, while the QFI is equal to the largest FI optimizing over all the possible measurements. In this paper the optimal projective measurement is constituted by the eigenvectors $|\psi_{m}\rangle$ of the final state, and the measured probabilities are the eigenvalues of the final state.
According to Eq.~(\ref{QFI}), it is easy to obtain the QFI in the estimation of the LIV parameter which is given by
\begin{eqnarray}
H(\alpha^2)&=&\sum_{n=0}^\infty\frac{(\partial_{\alpha^2}\lambda_n)^2}{\lambda_n},
\end{eqnarray}
where $\lambda_n=\gamma^n-\gamma^{n+1}$ is the eigenvalues of the diagonal density matrix.
Now let us study the behavior of the precision of the LIV parameter estimation which is investigated by the QFI which is dependent on $\alpha^2$, $k$, $\epsilon$ and $\rho$.

\begin{figure}[htbp]
\centering
\includegraphics[height=1.8in,width=2.8in]{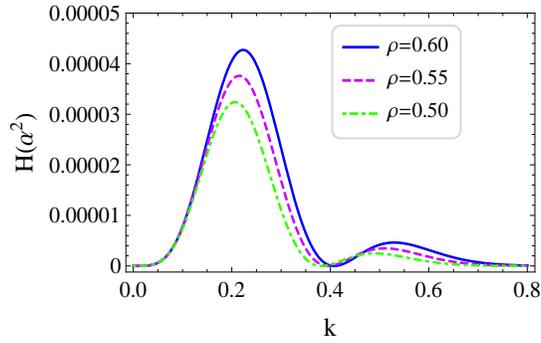}
\caption{ QFI in the estimation of the LIV parameter $\alpha^2$ as a function of momentum mode $k$ for the fixed values $\alpha^2=-10$, $\epsilon=0.99$, and three different values of expansion rate, i.e., $\rho=0.6$ (solid line), $\rho=0.5$ (dashed line), and $\rho=0.4$ (dotted-dashed line).}\label{QFI1}
\end{figure}

In Fig.~\ref{QFI1}, we plot the QFI in the estimation of the LIV parameter as a function of the momentum modes $k$ of the particles with different expansion rate $\rho$, for the fixed expansion rate $\epsilon=0.99$ and the LIV coefficient $\alpha^2=-10$.
We can see from the plot that in the extreme case of $k\rightarrow0$, i.e., at the large wavelengths (large spatial scales), the wavelengths reach to the classical length scale and we can not observe the effects of the LIV for the small momentum $k$ of the particles, which means that the QFI should tend to zero. On the other hand, in the extreme case of $k\rightarrow\infty$, i.e., at the small wavelengths (small spatial scales), the QFI vanishes because  the system approaches to the continuous limit where the effects of discreteness are not felt by the bosons. This means that when $k\rightarrow0$ and $k\rightarrow\infty$, the quantum parameter estimation reaches to the lowest precision.
Compared to that of quantum entanglement for a massless scalar field due to LIV during the spacetime expansion which is discussed in Ref. \cite{Khajeh}, we can conclude that the precision of LIV parameter estimation can be enhanced if entangled states are used as probe systems. However, it is worth noting that for a finite momentum modes $k$ of the particles, when the frequency of asymptotically flat regions in the past and future times satisfies
\begin{eqnarray}
\omega_{\rm in}\sinh(\frac{\pi}{\rho}\omega_{\rm in})=(1+2\epsilon)\omega_{\rm out}\sinh(\frac{\pi}{\rho}\omega_{\rm out}),
\end{eqnarray}
we get that the QFI equals to zero, which means that the QFI have two peaks as the increases of $k$. While we are interested in that there is a optimal momentum for each mode, $k_{\rm max}$, so that the QFI is maximum at this momentum mode, which indicates that the momentum modes $k$ have a range providing us a better precision.

\begin{figure}[htbp]
\centering
\includegraphics[height=1.8in,width=2.8in]{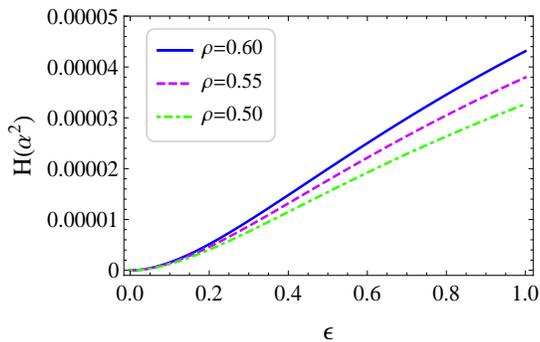}
\caption{ The maximum QFI in the estimation of the LIV parameter $\alpha^2$ as a function of expansion volume $\epsilon$, for the fixed values $\alpha^2=-10$ and three different values of $k$ and $\rho$, i.e., $k_{\rm max}$ with $\rho=0.6$ (solid line), $\rho=0.5$ (dashed line) and $\rho=0.4$ (dotted-dashed line).}\label{picture2}
\end{figure}
Figure \ref{picture2} shows that the maximum value of the QFI in the estimation of the LIV parameter obtained for optimal momentum mode $k$ varies with the expansion volume $\epsilon$, for the fixed LIV parameter $\alpha^2=-10$. Three different values of expansion rate $\rho$ have been discussed in this figure. We find that as the expansion rate grows, the precision of estimation of the LIV parameter increases accordingly. Moreover, it is obvious that the maximum QFI for the privileged value of $k$ increases by increasing the expansion volume $\epsilon$. This implies that we can get the highest precision of the LIV parameter estimation by choosing a larger value of expansion volume $\epsilon$.

\begin{figure}[htbp]
\centering
\includegraphics[height=1.8in,width=2.8in]{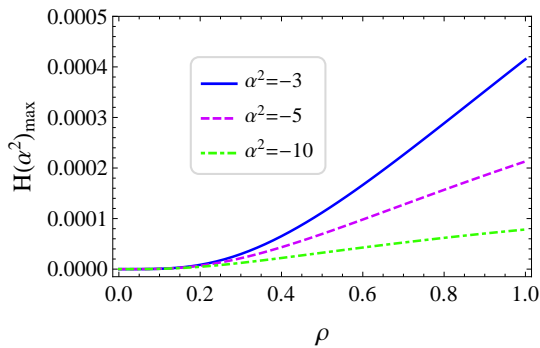}
\caption{ The maximum QFI in the estimation of the LIV parameter as a function of the expansion rate $\rho$ for the optimal momentum mode $k$, with fixed values $\epsilon=0.99$ and three different values of LIV parameter, i.e., $\alpha^2=-3$ (solid line), $\alpha^2=-5$ (dashed line), $\alpha^2=-10$ (dotted-dashed line).}\label{picture3}
\end{figure}

Figure \ref{picture3} presents the maximum QFI in term of the expansion rate $\rho$ for the optimal value of momentum mode $k$, with fixed value of the expansion volume $\epsilon=0.99$. Three different values of LIV parameter $\alpha^2$ have been shown in this plot. From this figure we can deduce that the QFI grows as the expansion rate gets larger values, which implies that the highest precision of estimation can be improved by a larger expansion rate $\rho$.
Moreover, the higher the LIV parameter $\alpha^2$, the bigger the QFI
is, i.e., the easier it is to achieve a given precision in the estimation of LIV parameter.

To sum up, we can choose larger cosmological parameters which are allowed for the expanding universe, as well as some suitable momentum modes of the particles to obtain the optimal strategy realizing the ultimate limit of precision in the estimation of the LIV parameter, such that we can easier to reach the highest precision for higher LIV parameter.

\section{Conclusions}

In this paper, we have studied the quantum parameters estimation for a massive scalar field on a two-dimensional asymptotically flat Robertson-Walker expanding spacetime in the presence of LIV.
Based on the theory of local quantum estimation, we have derived the ultimate bounds to precision in the cosmological parameters and LIV coefficient estimation, which is related to the QFI.
We have choose the entangled states which are used as probe system so as to improve the precision of the estimation.
In particular, by performing projective measurement on the eigenvectors of the particle states, the maximum FI is equal to the maximum QFI, which implies that this measurement is the optimal measurement for the estimation of cosmological and LIV parameters corresponding to the population measurement.
Moreover, it was shown that we can achieve the largest QFI by adjusting suitable LIV, cosmological and field parameters.
Two different situations, in the estimation of cosmological and LIV parameters, have been considered, respectively.

In the estimation of cosmological parameters, we noted that the bounds on measurement for the Lorentz-invariant massive scalar field case can be enhanced under some certain conditions due to the effects of LIV during the evolution of expansion spacetime. Since particles can be produced for the Lorentz-invariant massive scalar field and massless scalar field in the presence of LIV, we discussed two background structure cases
and shown that the optimal precision of estimation can be obtained by choosing appropriate LIV, cosmological and field parameters.

On the other hand, in the estimation of LIV parameter, we found that there are a range of momentum modes of the field particles that provide us a better precision in the estimation process. Furthermore, we should choose a larger value of cosmological parameters to enhance the precision for the estimation of the LIV parameter. In particular, as the LIV parameter gets larger values, it is easier to obtain the given precision in the estimation of LIV parameter.

%%%%%%%%%%%%%%%%%%%%%%%%%%%%%%%
\begin{acknowledgments}
This work was supported by the  National Natural Science Foundation
of China under Grant Nos. 11875025, 11475061 and 11675052.
Hunan Provincial Natural Science Foundation of China under Grant No. 2018JJ1016.
The project was funded by the CAS Key Laboratory for Research in Galaxies and Cosmology, Chinese Academy of Science (No. 18010203).

\end{acknowledgments}

\end{document}